\documentclass[3p,times,procedia]{elsarticle}
\usepackage{nupha_ecrc}

\volume{00}

\firstpage{1}

\journalname{Nuclear Physics A}

\runauth{S. McDonald, C. Shen, F. Fillion-Gourdeau, S.Jeon, C. Gale}

\jid{nupha}

\jnltitlelogo{Nuclear Physics A}

\usepackage{amssymb}
\usepackage{amsmath}

\usepackage[utf8]{inputenc}

\usepackage[figuresright]{rotating}
\usepackage{titlesec}
\titleformat{\subsection}[runin]{\bfseries}{}{0pt}{}[]

\begin{document}

\begin{frontmatter}



\dochead{XXVIth International Conference on Ultrarelativistic Nucleus-Nucleus Collisions\\ (Quark Matter 2017)}

\title{A Detailed Study and Synthesis of Flow Observables in the IP-Glasma+MUSIC+UrQMD Framework}


\author[McGill]{Scott McDonald}

\author[McGill,Brookhaven]{Chun Shen}
 
\author[INRS,Waterloo]{Fran\c{c}ois Fillion-Gourdeau}

\author[McGill]{Sangyong Jeon}
 
 \author[McGill]{Charles Gale}
 
 \address[McGill]{Department of Physics, McGill University, 3600 University
 Street, Montreal,
 QC, H3A 2T8, Canada}
 
 \address[Brookhaven]{Physics Department, Brookhaven National Laboratory, Upton, NY 11973, USA}
 
\address[INRS]{Universit\'e du Qu\'ebec, INRS-\'Energie, Mat\'eriaux et
T\'el\'ecommunications, Varennes, Qu\'ebec, Canada J3X 1S2}

\address[Waterloo]{Institute for Quantum Computing, University of Waterloo, Waterloo, Ontario, Canada, N2L 3G1}

\begin{abstract}
In this work we use the IP-Glasma+MUSIC+UrQMD framework to systematically study a wide range of hadronic flow observables at 2.76 TeV. In addition to the single particle spectra and anisotropic flow coefficients $v_n$ previously studied in \cite{1609.02958}, we consider event-plane correlations, non-linear response coefficients $\chi_{npq}$, and event shape engineering. Taken together, these observables provide a wealth of insight into the collective behavior of the QGP and initial state fluctuations. They shed light on flow fluctuations, flow at fixed system size but different initial geometries, as well as the non-linear hydrodynamic response to the initial state spatial eccentricities. By synthesizing this information we can gain further insight into the transport properties of the QGP as well as the fluctuation spectrum of the initial state.
\end{abstract}


\end{frontmatter}


\section{Introduction}
\label{Intro}
A number of models have been developed to study the formation and subsequent evolution of Quark Gluon Plasma (QGP), a deconfined state of quarks and gluons formed under extreme temperatures and pressures in heavy ion collisions at RHIC and the LHC. Many phenomenological studies of QGP investigate the sensitivity of one or more observables to a parameter or set of parameters within a model, with the goal of extracting information about QGP. The current study aims to take a more comprehensive look at flow observables within a single theoretical framework. Having used the hybrid IP-Glasma+MUSIC+UrQMD model to reproduce flow observables, the current work expands the study done in \cite{1609.02958} to consider more differential flow observables with the goal of further constraining the physics of the QGP.

\section{Model and Parameters} 
The IP-Glasma model, originally developed in \cite{Schenke:2012wb}, includes realistic event-by-event geometric and sub-nucleonic quantum fluctuations. A new implementation \cite{1609.02958} is used to initialize MUSIC \cite{Gale:2013da}, a second-order relativistic viscous hydrodynamics code. A constant shear viscosity to entropy density ratio is used $\eta/s=0.095$ along with a temperature dependent bulk viscosity $\zeta/s(T)$ based on \cite{NoronhaHostler:2008ju,Karsch:2007jc}. After hydrodynamic evolution, the fluid hadronizes at an isothermal hypersurface of $T_{sw}=145$ MeV, from which UrQMD is initialized. Hadrons then undergo resonance decays and hadronic re-scatterings in UrQMD \cite{nucl-th/9803035} before freezing out. A more in-depth discussion of the model, parameters, and centrality selection can be found \cite{1609.02958}. 

\section{Results}
It has been known for some time that IP-Glasma's initial state fluctuations are able to describe event-by-event distributions of $v_n$'s \cite{1209.6330} when coupled to viscous hydrodynamics. Such event-by-event flow fluctuations give rise to non-trivial flow correlations. 
By fixing the system size, and isolating fluctuations in the momentum space geometry in a given centrality bin, event shape engineering (ESE) gives a measure of the spread of values of $v_n$ while simultaneously illuminating correlations between different harmonics. In practice, ESE is done by re-binning centrality using the reduced flow vector \cite{Voloshin:2008dg},
\begin{equation}
\begin{split}
  q_n = \frac{Q_n}{\sqrt{N}}, \quad \quad 
  Q_n =\sum\limits_{i=1}^{N}{e^{in\phi_i},}
 \end{split}
\end{equation}
where $N$ is the number of particles in the event.
\begin{figure}[hb!]
\centering
      \includegraphics[width=0.7\textwidth]{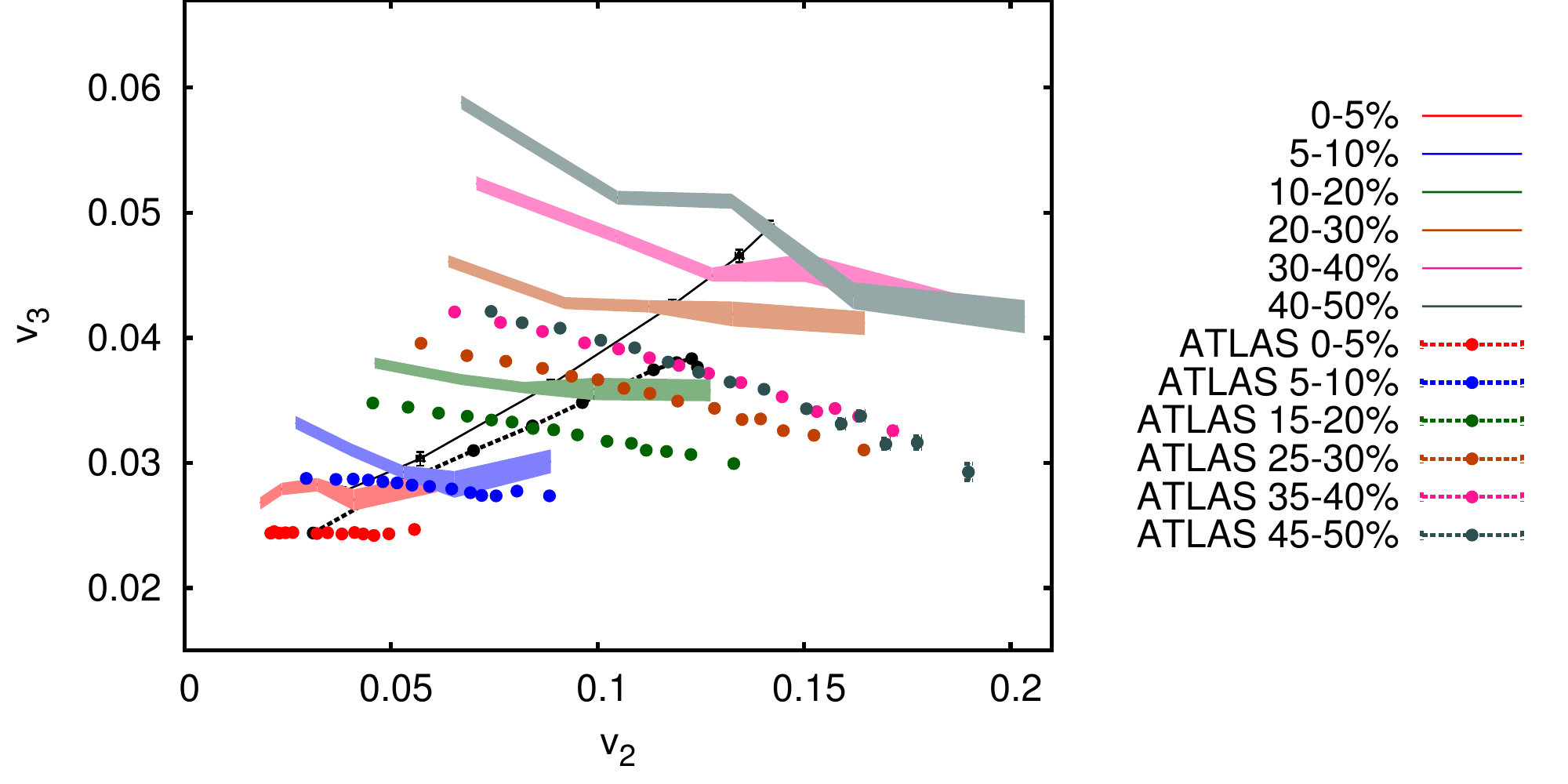}
       \caption{$v_3$ vs $v_2$ for $0.5 ~{\rm GeV} \le p_T \le 2.0 ~{\rm GeV}$ in $q_2$ bins at 2.76 TeV, compared to ATLAS data \cite{Aad:2015lwa}. The dashed black line represents the centrality bin averaged values from ATLAS for 0-50\%, and the solid black line corresponds to the same quantities from the IP-Glasma+MUSIC+UrQMD simulations.}
       \label{fig1}
\end{figure}
Fig.~\ref{fig1} shows Pb-Pb collisions at 2.76 TeV in $q_2$ bins.  Within each centrality bin, one can see a wide spread of values for $v_2$, and an anti-correlation between $v_2$ and $v_3$ for peripheral collisions. Such an anti-correlation has received much attention recently with the non-symmetric cumulant \cite{ALICE:2016kpq}, but is unmistakably present in ESE in both experiment and theory. It is worth noting that the overestimation of the data in Fig~.\ref{fig1} is largely due to the fact that our transport coefficients, $\eta/s$ and $\zeta/s(T)$, were extracted in \cite{1502.01675} from particle spectra and integrated $v_n$ at the expense of not describing the differential $v_n(p_T)$ for all $p_T$. This was done with ALICE data for which $0.2 ~{\rm GeV} \le p_T \le 5.0 ~{\rm GeV}$. Considering the $p_T$-dependence of $v_n(p_T)$ in \cite{1609.02958}, where we slightly underestimate the data below about 0.7 GeV and overestimate above this point, it is clear that the ATLAS $p_T$-cut will lead to overestimates of the integrated flow harmonics. Nonetheless, the spread of values and the correlation between $v_2$ and $v_3$ that are present in the experimental data are qualitatively reproduced in the simulations. 

The ESE provides many data points by which to constrain model parameters, and the relationship between different centrality averaged $v_n$'s, sometimes referred to as the "boomerang," has been shown to be quite sensitive to the shear viscosity \cite{Qian:2016pau}. These features make ESE an excellent candidate to constrain the shear and bulk viscosities of the QGP through direct comparison between theory and experiment.

Event-plane correlators, on the other hand, provide a more direct way to quantify correlations between flow harmonics, and are defined for the two-plane correlator as  
\begin{equation}
\begin{split}
\cos\left[c_1 n_1 \Psi_{n_1} - c_2 n_2 \Psi_{n_2} \right] 
 = \frac{\Re\{\langle Q_{n_1}^{c_1} (Q_{n_2}^{c_2})^* \rangle \}}{\sqrt{\langle Q_{n_1}^{c_1} (Q_{n_1}^{c_1})^* \rangle} \sqrt{\langle (Q_{n_2}^{c_2}) (Q_{n_2}^{c_2})^* \rangle}} \quad 
 \hbox{with} \quad \sum\limits_{i=1}^{N} c_in_i=0.
\end{split}
\end{equation}
\begin{figure*}[hb!]
  \centering
  \begin{tabular}{ccc}
             \includegraphics[width=.35\textwidth]{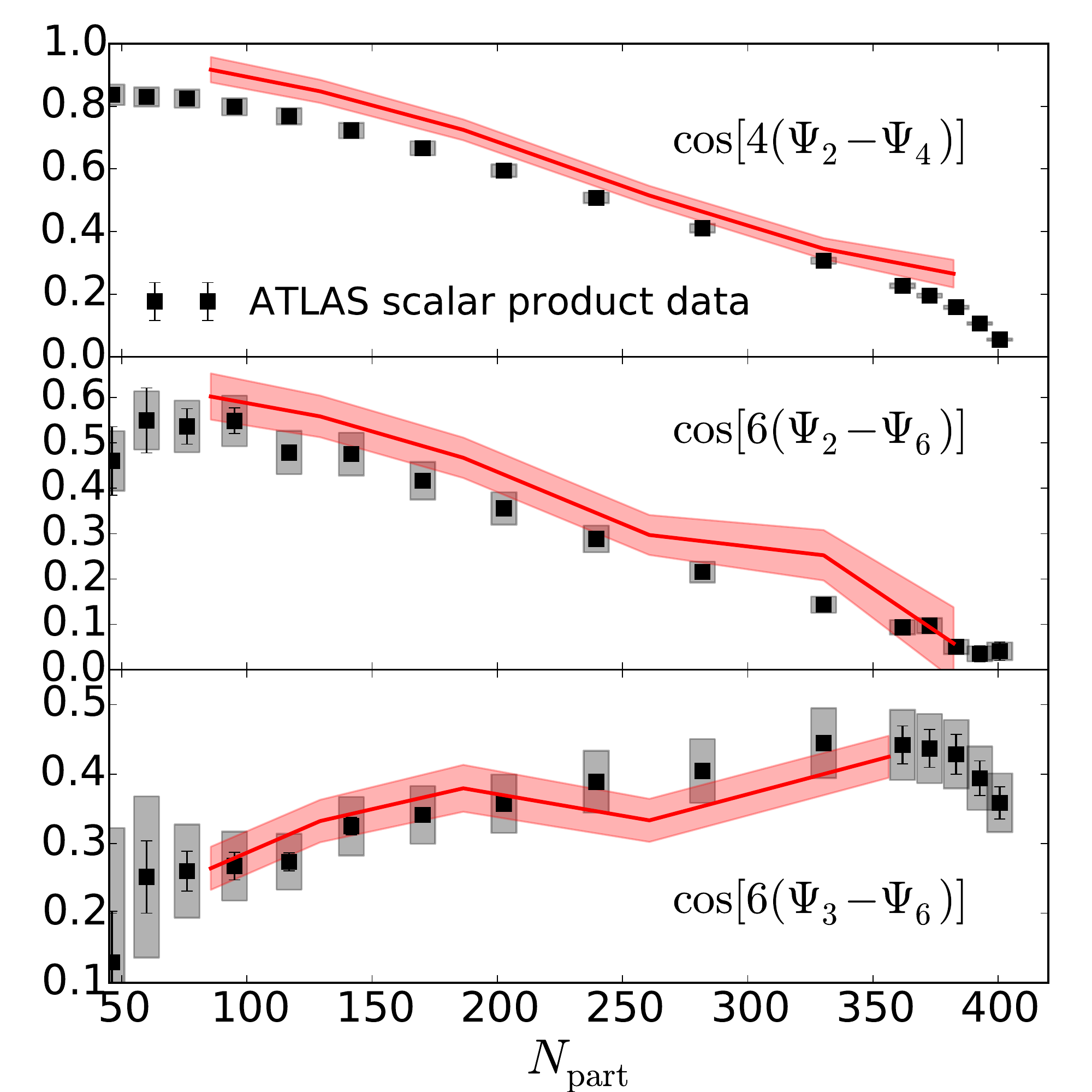}
        \includegraphics[width=.3\textwidth]{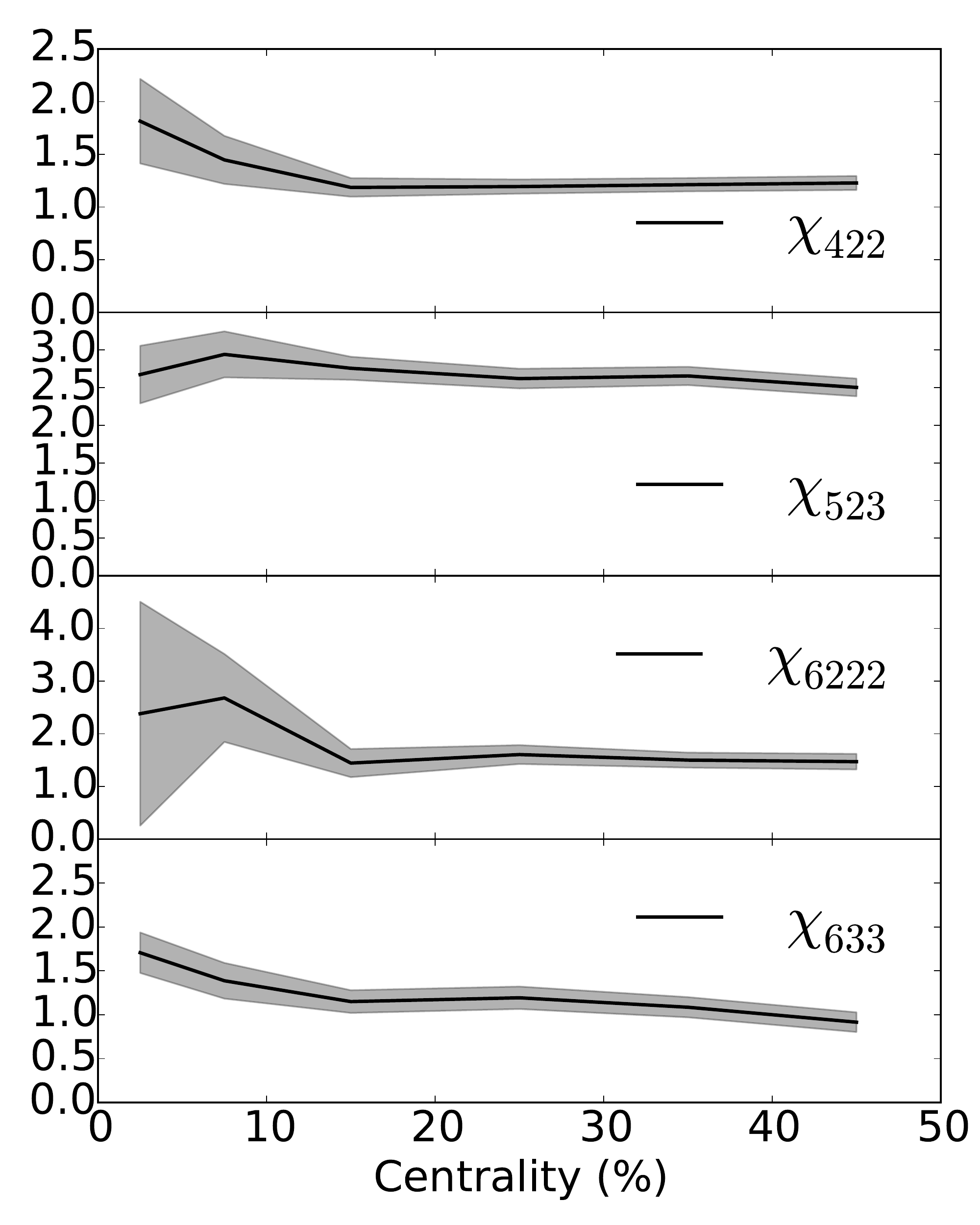} 
          \includegraphics[width=.35\textwidth]{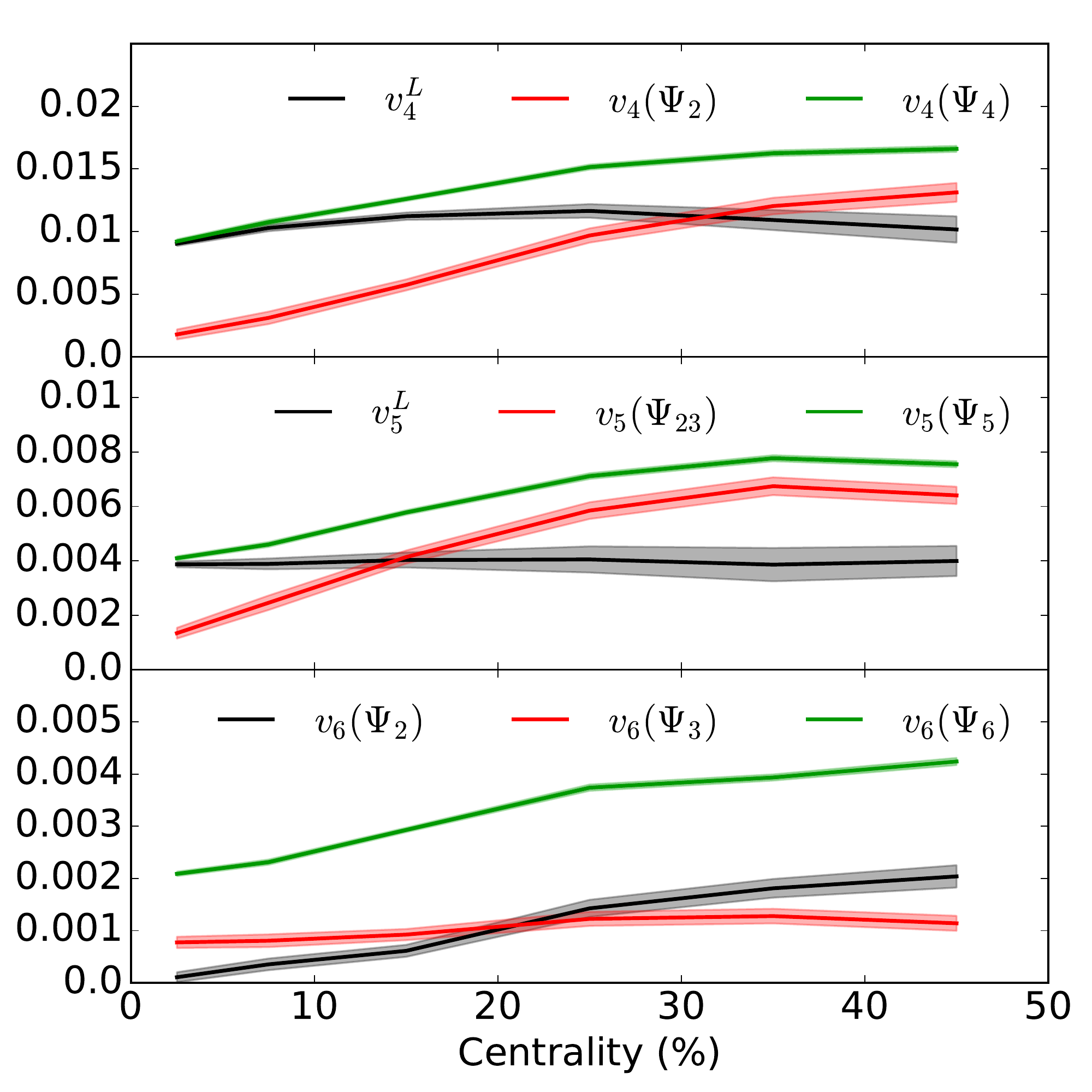}
\end{tabular}
 \caption{Left: Event-plane correlators as function of centrality vs. ATLAS data \cite{Aad:2014fla}. Center: Predictions for non-linear response coefficients $\chi_n$ as a function of centrality. Right: Predictions for Linear and non-linear response terms plotted along with the full harmonics, for $v_4$, $v_5$, and $v_6$. Note $v_6^L$ is not plotted because it is approximately two orders of magnitude smaller than $v_6\{\Psi_2\}$ and $v_6\{\Psi_3\}$. All three panels are for Pb-Pb collisions at 2.76 TeV. }
 \label{fig2}
\end{figure*} 
Three such event plane combinations are plotted in the left panel of Fig.~\ref{fig2} The event-plane correlators can be understand in terms of the formalism that decomposes the Fourier coefficients $v_n$ into a linear and non-linear response. 
One can write each $v_n$ as a series expansion in the initial state energy anisotropies $\mathcal{E}_n $, 
\begin{equation}
    V_n = \kappa_n \mathcal{E}_n +\sum\limits_{n=p+q}{\kappa_{npq}^{'}\mathcal{E} _{p}\mathcal{E}_{q}}+... = V_n^L+\sum\limits_{n=p+q}{\chi_{npq}V_{p}V_{q}+...}
\end{equation}
where $V_n^L$ is the linear response term that is proportional to $\mathcal{E}_n$, and $V_p$, $V_q$ are lower order modes that couple to contribute to $V_n$. The $\mathcal{E}_n$ can be considered as input into the hydrodynamic evolution, and the coefficients $\kappa_n$, $\kappa_{npq}^{'}$, and $\chi_{npq}$, as functions of QGP transport properties that characterize the hydrodynamic response to the initial state. 
Consider the decomposition of $v_6$, given in \cite{Yan:2015jma} as
\begin{equation}\label{eq:v6}
V_6=V_{6}^L+\chi_{6222}(V_2)^3 +\chi_{633}(V_3)^2 = V_{6}^L+ V_6\{\Psi_2\} +V_6\{\Psi_3\},
\end{equation}
\begin{equation}
\begin{split}
\chi_{6222}= \frac{\langle V_6(V_2^*)^3 \rangle}{\langle |V_2|^6 \rangle}, \quad \quad
\chi_{633}= \frac{\langle V_6(V_3^*)^2 \rangle}{\langle |V_3|^4 \rangle}.
\end{split} 
\end{equation}
Predictions for the $\chi_{npq}$ coefficients and the $v_n$ decomposition for $n=4,5,6$ are plotted in the central and right panels of Fig.~\ref{fig2}. Note that upper-case $V_n$ represents the single event flow vector, whereas lower-case $v_n$ is the RMS value. The RMS value of the linear response is calculated as $v_n^L=\sqrt{v_n^2-(\chi_{npq}v_pv_q)^2}$.

The $\psi_6$-$\psi_2$ and $\psi_6$-$\psi_3$ correlators can be understood in terms of the decomposition in Eq.~\ref{eq:v6}. As one moves from central to peripheral collisions, the correlation between the $\psi_6$ and $\psi_2$ event planes increases as the global geometry leads to an increase in the contribution from $v_2$. The increasing contribution from $v_2$ leads to a diminished contribution from $v_3$, and thus a decreasing correlation between $\psi_6$ and $\psi_3$. This can be seen quite clearly in the right panel of Fig.~\ref{fig2}, where $v_6\{\Psi_{2}\}$ overtakes $v_6\{\Psi_3\}$ at about 20\% centrality.  
\begin{figure}
\centering
       \includegraphics[width=.5\textwidth]{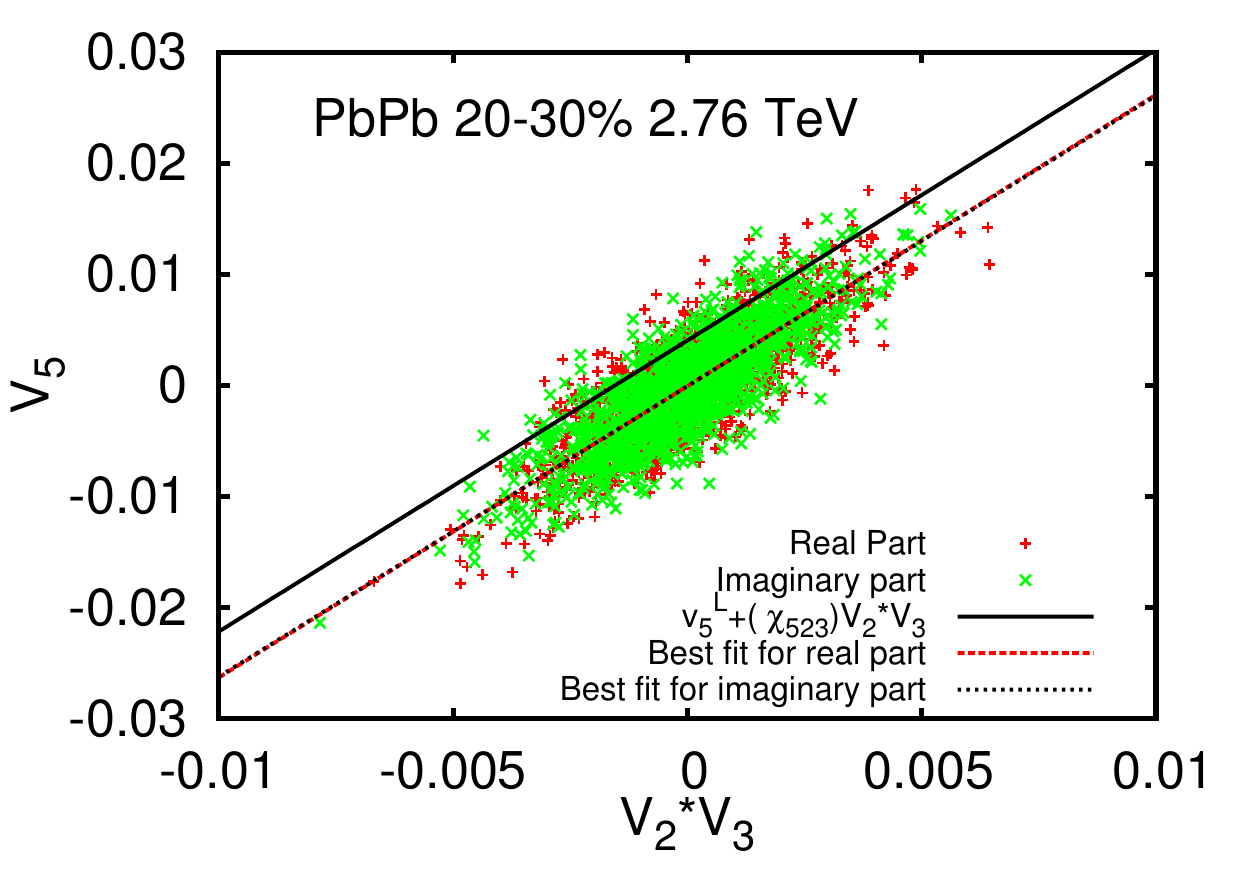}
       \caption{A scatter plot of the real and imaginary parts of $V_5$ vs. $V_2V_3$ of individual IP-Glasma+MUSIC+UrQMD events, along with the best fit lines for both. Plotted in black is the calculated value for $V_5=v_5^L+\chi_{523}V_2V_3$.}
       \label{fig3}
\end{figure}

To verify that the linear and non-linear response terms are describing the event-by-event calculations, it is instructive to compare the expressions with a scatter plot of individual events from the simulations. In Fig.~\ref{fig3} the real and imaginary parts of $V_5$ have been plotted as a function of $V_2 V_3$, along with best fit lines. It is clear that the line $V_5 = v_5^L + \chi_{523}V_2V_3$ is very nearly parallel to both of the best fit lines. This means that $\chi_{523}$ is indeed the slope of $V_5$ plotted as a function of $V_2V_3$, and the non-linear response formalism is working. In the vertical direction, the spread of values around the best fit line can be thought of as the range of $V_5^L$, which fluctuates independently of $V_2V_3$ on an event-by-event basis. The calculated $v_5^L$ is the RMS value, so it is to be expected that it overestimates the y-intercepts of the best fit lines.  
\section{Conclusion} 
 In this work, we have expanded the study in \cite{1609.02958} to include event shape engineering as well as the linear and non-linear mode-coupling terms of the $v_n$ flow harmonics. We offer an explanation of the event-plane correlators in terms of the non-linear response terms, while making predictions for the latter observables. Finally, it was shown that the non-linear decomposition of the flow harmonics can describe event-by-event simulations.

\subsection{Acknowledgements}
This work was supported in part by the Natural Sciences and Engineering Research Council of Canada, as well as the U.S. Department of Energy, Office of Science under contract No. DE- SC0012704. Computations were made in part on the supercomputer Guillimin from McGill University, managed by Calcul Qu\'ebec and Compute Canada. The operation of this supercomputer is funded by the Canada Foundation for Innovation (CFI), NanoQu\'ebec, RMGA and the Fonds de recherche du Qu\'ebec - Nature et technologies (FRQ-NT). C. G. gratefully acknowledges support from the Canada Council for the Arts  through its Killam Research Fellowship program. C.S. gratefully acknowledges a Goldhaber Distinguished Fellowship from Brookhaven Science Associates.





\bibliographystyle{elsarticle-num}







\end{document}